\documentclass[letter]{jpsj3}

\title{Superconductivity in Pseudo-Binary Silicide SrNi$_x$Si$_{2-x}$ \\ with AlB$_2$-Type Structure}

\author{
Sunseng \textsc{Pyon}$^{1,2}$\thanks{E-mail: pyon@science.okayama-u.ac.jp}, Kazutaka \textsc{Kudo}$^{1,2}$, and Minoru \textsc{Nohara}$^{1,2}$
}
\inst{\address{2-31-22-5F Yushima, Bunkyo-ku, Tokyo 113-0034}
}

\inst{$^1$Department of Physics, Faculty of Science, Okayama University, Okayama  700-8530, Japan \\
$^2$Transformative Research-Project on Iron Pnictides (TRIP), Japan Science and Technology Agency (JST), 5 Sanbancho, Chiyoda-ku, Tokyo 102-0075, Japan}

\abst{
We demonstrate the emergence of superconductivity in pseudo-binary silicide SrNi$_x$Si$_{2-x}$.
The compound exhibits a structural phase transition from the cubic SrSi$_2$-type structure ($P4_132$) to the hexagonal AlB$_2$-type structure ($P6/mmm$) upon substituting Ni for Si at approximately $x = 0.1$.
The hexagonal structure is stabilized in the range of $0.1 < x <$ 0.7.
The superconducting phase appears in the vicinity of the structural phase boundary.
Ni acts as a nonmagnetic dopant, as confirmed by the Pauli paramagnetic behavior.
}

\kword{superconductivity, alkaline-earth disilicide, SrSi$_2$, AlB$_2$-type structure}

\begin{document}

\maketitle
Alkaline-earth disilicides AESi$_2$ (AE = Ca, Sr, or Ba) exhibit a rich polymorphism owing to their lattice instabilities.
Six polymorphic forms have been identified in AESi$_2$: orthorhombic BaSi$_2$-type ($Pnma$), cubic SrSi$_2$-type ($P4_132$), trigonal EuGe$_2$-type ($P\bar{3}m1$), tetragonal $\alpha$-ThSi$_2$-type ($I4_1/amd$), trigonal CaSi$_2$-type ($R\bar{3}m1$), and hexagonal AlB$_2$-type ($P6/mmm$)\cite{ref:Imai1, ref:Evers1, ref:Evers2, ref:Bordet, ref:Evers3}.
Binary SrSi$_2$ shows dimorphism, with a cubic SrSi$_2$-type and a tetragonal $\alpha$-ThSi$_2$-type.
The cubic structure transforms into the tetragonal one under hydrostatic pressures or at elevated temperatures\cite{ref:Evers1}.
Cubic SrSi$_2$ is a narrow-gap semiconductor\cite{ref:Imai2,ref:McWhan},
whereas tetragonal SrSi$_2$ is a metal and exhibits superconductivity below 3.1 K\cite{ref:Evers4}.
Interestingly, the hexagonal AlB$_2$-type structure can be stabilized when foreign elements, such as Al and Ga, are partially substituted for Si in SrSi$_2$. Examples are the pseudo-binary compounds Sr(Al,Si)$_2$\cite{ref:Lorenz, ref:Evans} and Sr(Ga,Si)$_2$\cite{ref:Imai3, ref:Meng}, which crystallize into the AlB$_2$-type hexagonal structure.
These compounds exhibit superconductivity at a superconducting transition temperature of $T_c$ = 4.7 and 5.1 K, respectively.
It is widely believed that lattice instabilities often lead to high-$T_c$ superconductivity.
The AESi$_2$ system can be a candidate for exploring higher $T_c$ values because their rich polymorphism suggests the presence of strong lattice instabilities\cite{ref:Lorenz, ref:Evans, ref:Imai3, ref:Meng, ref:Imai4, ref:Sung, ref:Imai5, ref:Hor, ref:Imai6, ref:Yamanaka}.

In this Letter, we report superconductivity in pseudo-binary silicide SrNi$_x$Si$_{2-x}$ with the AlB$_2$-type structure.
When Ni is partially substituted for Si in SrSi$_2$, the structure changes from the cubic SrSi$_2$-type structure into the hexagonal AlB$_2$-type structure\cite{ref:Nasir}.
We demonstrate that the hexagonal structure is stable over a wide range of $0.1 < x <$ 0.7 in SrNi$_x$Si$_{2-x}$. The system exhibits Pauli paramagnetic behavior even though magnetic element Ni is substituted.
Furthermore, superconductivity appears at the cubic-hexagonal phase boundary.

Polycrystalline SrNi$_x$Si$_{2-x}$ samples were prepared by arc-melting.
Mixtures with a  Sr:Ni:Si ratio of 1:$x$:$(2-x)$ (where $x =$ 0.1, 0.2, 0.3, 0.4, 0.5, 0.6, 0.7, 0.8, and 1.0) were pelletized and melted in an Ar atmosphere using an arc furnace.
The structure of the samples was examined by powder X-ray diffraction (XRD) measurements.
Magnetization was measured using a SQUID magnetometer (Quantum Design MPMS).
Electrical resistivity was measured by a standard DC four-terminal method using a Quantum Design PPMS.
Specific heat was measured by a relaxation method using a PPMS.

XRD measurements clarified that a single phase of polycrystalline samples was successfully synthesized for $x =$ 0.2, 0.3, 0.4, 0.5, and 0.6.
In the case of $x$ = 0.1, the cubic phase (the SrSi$_2$-type structure) was obtained with a small amount of the hexagonal phase (the AlB$_2$-type structure).
The $x$ = 0.7 sample was a mixture of the hexagonal phase and impurity phase(s).
The hexagonal phase was not obtained for $x \geq$ 0.8, suggesting a solubility limit at approximately $x$ = 0.7.
Lattice parameters $a$ and $c$ in the hexagonal phase at 0.1 $\leq$ $x$ $\leq$ 0.7, determined by XRD analysis, are shown in Fig. 1(a) as a function of $x$.
The $a$ parameter increases and the $c$ parameter decreases with increasing Ni content $x$.
The monotonic  behavior of the lattice parameters suggests that a solid solution of SrNi$_x$Si$_{2-x}$ was successfully obtained in the hexagonal phase.
The evolution of the crystal structures as a function of $x$ is summarized in Fig. 1(b).
The structural phase boundary between the cubic and hexagonal phases exists at approximately $x =$ 0.1.
The hexagonal structure in SrNi$_x$Si$_{2-x}$ is stabilized in the range  $0.1 < x <$ 0.7.
The solubility limit is located at $x$ $\simeq$ 0.7.

\begin{figure}[t]
\includegraphics[width=0.87\linewidth]{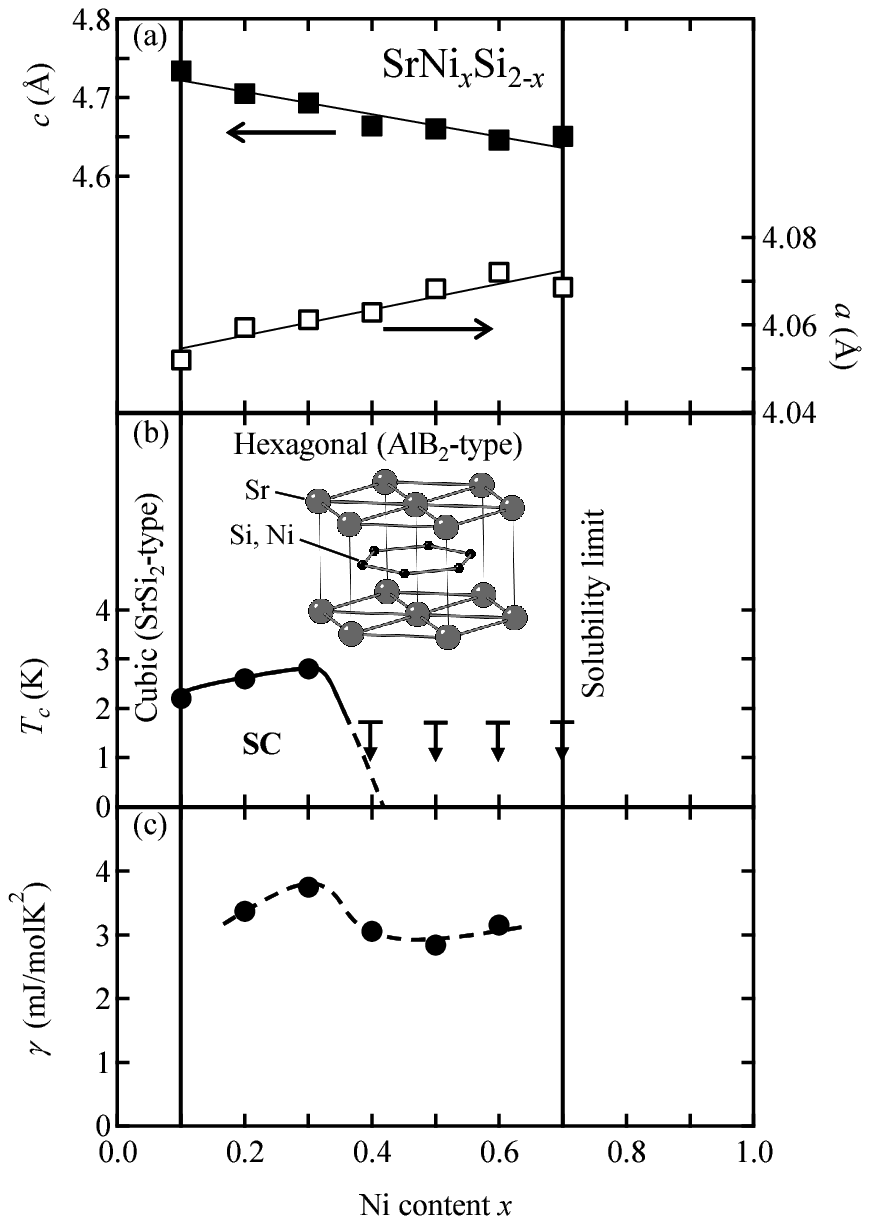} 
\caption{\label{fig1}
(a) Lattice parameters, determined from XRD measurements, in the hexagonal lattice at room temperature as functions of $x$ for SrNi$_x$Si$_{2-x}$.
(b) Structural and superconducting phase diagram of SrNi$_x$Si$_{2-x}$.
Closed circles indicate superconducting transition temperatures $T_c$ determined from magnetization measurements.
Bars and arrows indicate the absence of bulk superconductivity above 1.8 K.
The inset schematically illustrates the crystal structure of hexagonal AlB$_2$-type SrNi$_x$Si$_{2-x}$.
(c) Electronic specific-heat coefficient $\gamma$ as a function of $x$ for SrNi$_x$Si$_{2-x}$, which was estimated from linear extrapolation of the $C/T$-versus-$T^2$ data in the normal state.
}
\end{figure}

\begin{figure}[t]
\includegraphics[width=1\linewidth]{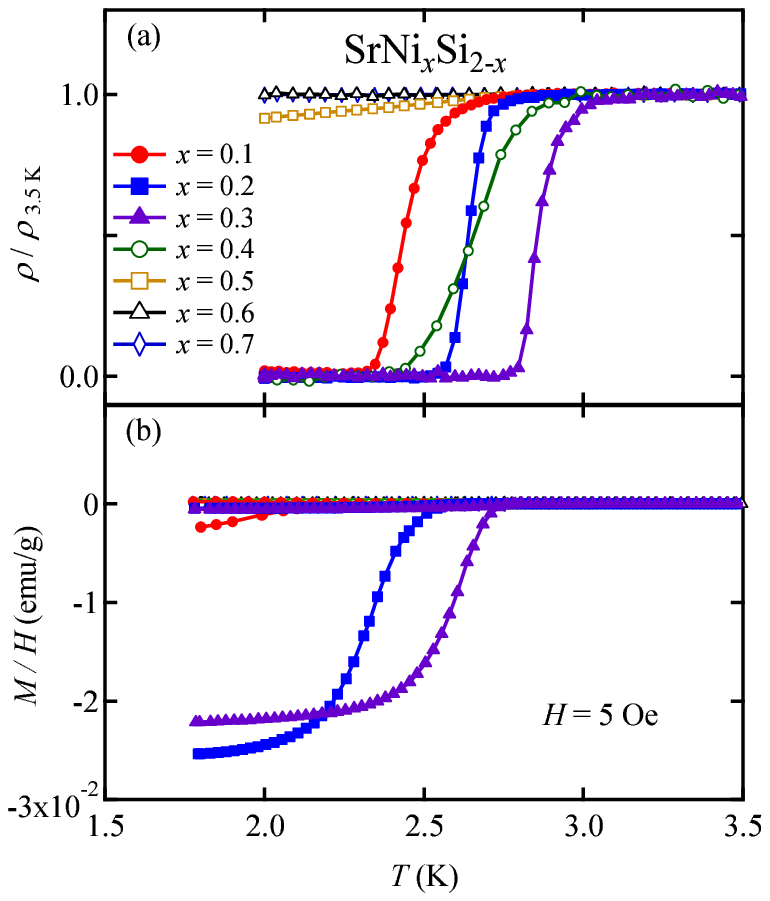}
\caption{\label{fig2}
(Color online) 
(a) Low-temperature resistivity of SrNi$_x$Si$_{2-x}$ (normalized by the value of $\rho$ at $T$ = 3.5 K) in zero applied field.
(b) Low-temperature magnetization of SrNi$_x$Si$_{2-x}$. The measurements of magnetization were conducted in an applied field of $H$ = 5 Oe with zero-field-cooled and field-cooled processes.
The shielding volume fraction for $x =$ 0.2 and 0.3 reached 119\% and 106\%, respectively, at 1.8 K.
}
\end{figure}

In this system, SrNi$_x$Si$_{2-x}$, a superconducting phase emerges in a narrow Ni-doping range near the cubic-hexagonal boundary at the hexagonal side, as shown in Fig. 1(b).
This is illustrated by the low-temperature resistivity $\rho$ and the magnetization $M$ data shown in Figs. 2(a) and 2(b), respectively.
For $x$ = 0.1, 0.2, and 0.3, zero resistivities and diamagnetic signals, a hallmark of superconductivity, were detected below $T_c$ = 2.3, 2.6, and 2.8 K, respectively.
The shielding volume fractions were 119\% and 106\% for the $x =$ 0.2 and 0.3 samples, respectively.
The shielding volume fraction was significantly lower for the $x = 0.1$ sample, because this sample is dominated by the nonsuperconducting cubic phase; the superconductivity at $T_c$ = 2.3 K can be ascribed to the hexagonal phase at $x$ = 0.1.
Bulk superconductivity was not detected for the $x =$ 0.4, 0.5, 0.6, or 0.7 sample.
Although the $x =$ 0.4 sample showed zero resistivity, a negligibly small diamagnetic signal (with a shielding volume fraction lower than 0.1\% at 1.8 K) suggests that the superconducting phase was not a bulk one for $x =$ 0.4.
For the $x =$ 0.5, 0.6 and 0.7 samples, neither zero resistivity nor diamagnetic behavior was observed above 1.8 K.

For further understanding of the superconducting state in SrNi$_x$Si$_{2-x}$, we performed resistivity measurements in applied magnetic fields to determine the upper critical field $H_{c2}$ of the $x$ = 0.3 sample, at which the highest $T_c$ was observed.
Figure 3(a) shows the temperature dependence of $H_{c2}$, which was determined from the midpoint of the resistive transition, shown in Fig. 3(b).
The extrapolation to 0 K using the Werthamer-Helfand-Hohenberg (WHH) theory \cite{ref:Werthamer} gives an estimate for the $H_{c2}(0)$ value of 1.58 T.
The $H_{c2}(0)$ value was roughly similar to that of a pseudo-binary alkaline-earth silicide Ca(Al,Si)$_2$ with an AlB$_2$-type structure\cite{ref:Ghosh, ref:Tamegai}.
We estimated the coherence length to be $\xi_0$ = $(\Phi_0/[2\pi H_{c2}(0)])^{1/2}$ = 144 \AA, where $\Phi_0$ is the magnetic flux quantum.
The Pauli limiting field $H_{\rm Pauli} = 1.84  T_c = 5.2$ T was larger than the estimated $H_{c2}$.
These results demonstrate that SrNi$_x$Si$_{2-x}$ is a conventional type-II superconductor.

To understand the role of Ni, we compared the normal-state resistivity of the superconducting sample ($x =$ 0.2) with that of the nonsuperconducting sample ($x =$ 0.6).
Figure 4(a) shows  $\rho$ as a function of temperature from 2 to 300 K for the $x =$ 0.2 and 0.6 samples.
$\rho$ exhibits an almost temperature-independent behavior above $T_c =$ 2.6 K for $x =$ 0.2, suggesting that the sample is a bad metal.
In contrast to that, $\rho$ for $x =$ 0.6 shows a noticeable metallic behavior with a positive temperature coefficient.
Another noticeable feature is the reduction of the residual resistivity for the $x$ = 0.6 sample.
We naively expect that disorder increases with Ni doping $x$, resulting in an increase in the residual resistivity.
However, what we observed was a decrease in the residual resistivity with $x$.
Thus, the observed reduction of the residual resistivity suggests an increase in carrier concentration with  Ni doping.
In other words, Ni acts to dope charge carriers in SrNi$_x$Si$_{2-x}$.

Magnetic susceptibility as a function of temperature is shown in Fig. 4(b) for the superconducting and the nonsuperconducting samples from 10 to 300 K.
The data show almost temperature-independent Pauli paramagnetic behaviors.
This behavior is characteristic of metals and is consistent with the (metallic) behavior of the resistivity.
By assuming the core diamagnetism of Sr$^{2+}$ to be $-20 \times 10^{-6}$ emu/mol,
we estimate the Pauli paramagnetic susceptibility to be $\chi_{\rm Pauli}$ $\simeq$ $25  \times 10^{-6}$ and $\simeq 50  \times 10^{-6}$ emu/mol for $x$ = 0.2 and 0.6, respectively.
These values are quite small and are comparable with those of $sp$ metals, suggesting that the electronic density of state at the Fermi level is small and the contribution of Ni $3d$ orbitals is negligible even for the heavily Ni-doped sample ($x$ = 0.6).
This is consistent with the small electronic specific-heat coefficient $\gamma$ $\simeq$ 3.8 mJ/molK$^2$ (for $x =$ 0.3), as we describe in the following.

\begin{figure}[t]
\includegraphics[width=1\linewidth]{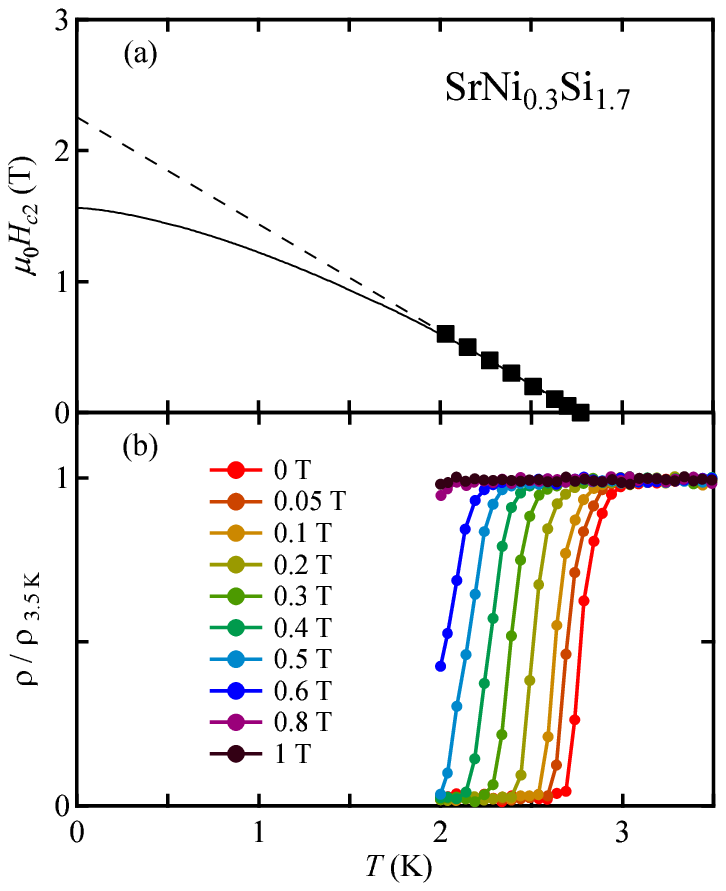}
\caption{\label{fig3}
(Color online) 
(a) Temperature dependence of the upper critical field $H_{c2}$ for SrNi$_x$Si$_{2-x}$ with $x$ = 0.3.
The solid line shows the WHH behavior. The broken line yields a slope of $-dH_{c2}/dT|_{T=T_c} = 0.82$ T/K.
(b) Low-temperature resistivity in various applied fields for SrNi$_x$Si$_{2-x}$ with $x$ = 0.3.
}
\end{figure}

The superconducting transition temperature $T_c$ increases monotonously with Ni doping at 0.1 $\leq$ $x$ $\leq$ 0.3 in SrNi$_x$Si$_{2-x}$.
Intriguingly, the superconducting phase disappears suddenly at approximately $x$ = 0.35, as shown in Fig. 1(b). 
This non-monotonous dependence of $T_c$ on doping $x$ accords with the electronic density of states (DOS) at the Fermi level, and thus with the electronic specific-heat coefficient $\gamma$.
We determined $\gamma$ from the specific heat $C$ using the $C/T$-versus-$T^2$ plot in Fig. 5, and plotted $\gamma$ as a function of $x$ in Fig. 1(c).
$\gamma$ increases with $x$ in the superconducting region (0.1 $<$ $x$ $\leq$ 0.3), at which $T_c$ increases with $x$.
$\gamma$ suddenly decreases at $x$ $\simeq$ 0.4, above which superconductivity disappears.
Such non-monotonous changes of $T_c$ on doping have been reported on CaAl$_{2-x}$Si$_x$\cite{ref:Lorenz} and SrGa$_x$Si$_{2-x}$ \cite{ref:Meng}.
Contrary to the decrease in the specific-heat coefficient $\gamma$ at $x$ $\simeq$ 0.4, the value of Pauli paramagnetic susceptibility increases; $\chi_{\rm Pauli}$ $\simeq$ 25 $\times$ 10$^{-6}$ emu/mol for the superconducting region, while $\chi_{\rm Pauli}$ $\sim$ 50 $\times$ 10$^{-6}$ emu/mol for the nonsuperconducting region at $x$ $\geq$ 0.4, as can be seen from Fig. 4(b). 
We do not understand the reason for this at present.
Further studies are invaluable. 

Specific heat shows a clear jump at $T_c$, indicative of bulk superconductivity, for $x =$ 0.2 and 0.3, as shown in Fig. 5.
We estimate the ratio of specific-heat jump at $T_c$ to $\gamma T_c$ to be $\Delta C(T_c)/\gamma T_c$ $\simeq$ 0.8, using $\gamma$ = 3.8 mJ/molK$^2$ and $\Delta C(T_c)/T_c$ = 3.2 mJ/molK$^2$ for $x = 0.3$.
The ratio is smaller than 1.43, the value of the BCS weak coupling limit.
We ascribe this to the presence of a residual normal-state region, likely due to the inhomogeneity of the samples.

\begin{figure}[t]
\includegraphics[width=1\linewidth]{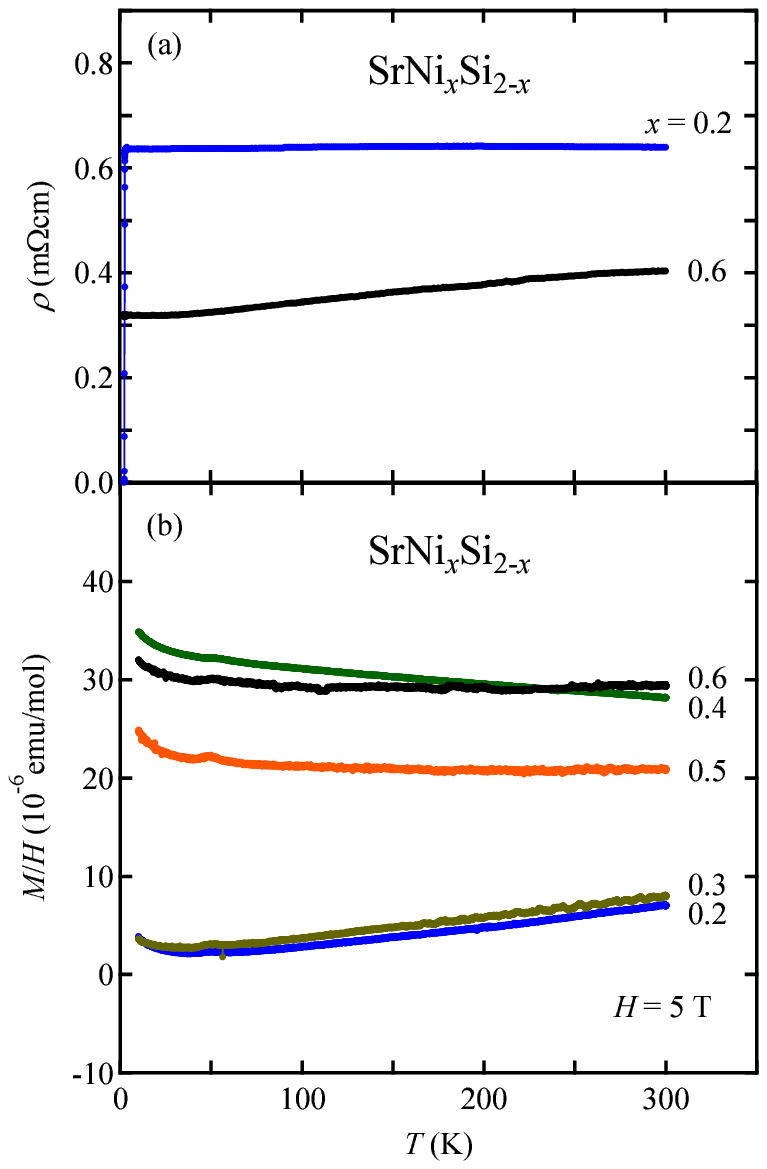}
\caption{\label{fig4}
(Color online) 
(a) Electrical resistivity $\rho$ of SrNi$_x$Si$_{2-x}$ for $x$ = 0.2 and 0.6, as a function of temperature.
(b) Temperature dependence of magnetization $M$ divided by $H$, $M/H$, in a magnetic field $H$ of 5 T for SrNi$_x$Si$_{2-x}$.
}
\end{figure}

Two intriguing features emerged from this study.
First, Ni acts as a nonmagnetic dopant; superconductivity emerged by substituting Ni for Si in SrNi$_x$Si$_{2-x}$.
In general, magnetic elements are harmful for superconductivity because of  severe magnetic pair breaking.
In contrast to this fate, the magnetic element Ni acts as a nonmagnetic dopant without inducing any local magnetic moments in  hexagonal SrNi$_x$Si$_{2-x}$ in the entire $x$ range of $0.1 < x <$ 0.7.
We naively expect that Ni $3d$ orbitals are completely filled in SrNi$_x$Si$_{2-x}$, as inferred from the small Pauli susceptibility.
This finding suggests that magnetic elements can be utilized as a dopant for producing superconductivity.
Second, the superconductivity of SrNi$_x$Si$_{2-x}$ emerged at the critical vicinity of the cubic-hexagonal phase boundary.
This result suggests that the structural instabilities play an important role in the occurrence of superconductivity in SrNi$_x$Si$_{2-x}$.
A similar phenomenon is known in Ba$_{1-x}$K$_x$BiO$_3$, in which superconductivity at up to $T_c$ = 32 K appears in the vicinity of the cubic and orthorhombic phase boundary\cite{ref:Cava, ref:Hinks, ref:Pei}.
The mechanism of superconductivity in Ba$_{1-x}$K$_x$BiO$_3$ is still in debate, though the importance of structural instabilities for describing the physical properties of Ba$_{1-x}$K$_x$BiO$_3$, including the origin of the superconductivity, is recognized \cite{ref:Zherlitsyn}.
The analogy between the two systems suggests the importance of lattice instabilities in the occurrence of the superconductivity in SrNi$_x$Si$_{2-x}$.

\begin{figure}[t]
\includegraphics[width=1\linewidth]{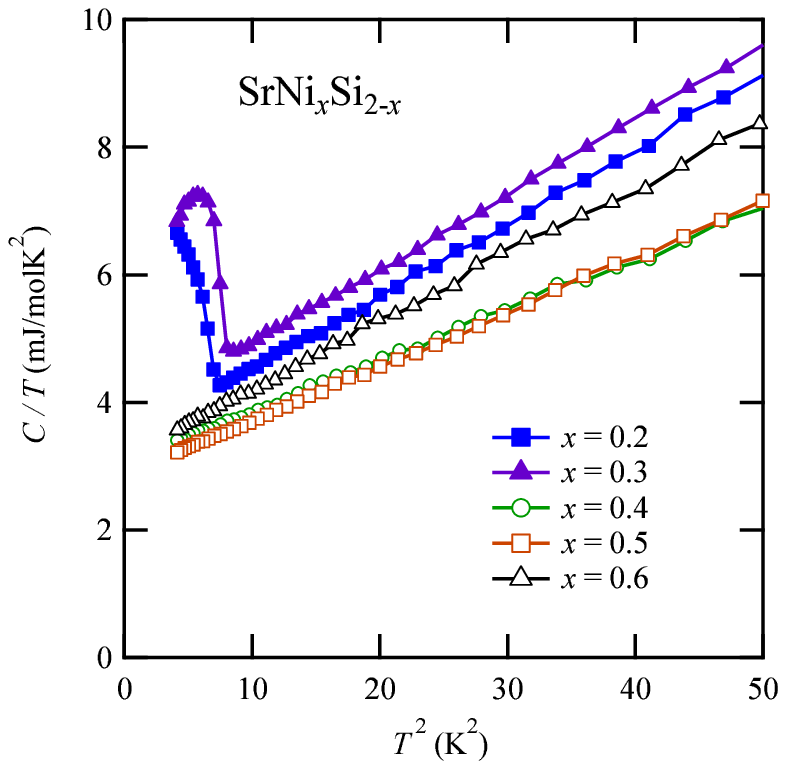}
\caption{\label{fig5}
(Color online) Specific heat divided by temperature, $C/T$, as a function of $T^2$ for SrNi$_x$Si$_{2-x}$ in zero applied field.
}
\end{figure}

In summary, pseudo-binary silicides SrNi$_x$Si$_{2-x}$ were synthesized by arc-melting, and their magnetic and electrical properties were characterized.
It was demonstrated that the semiconducting phase with a cubic SrSi$_2$-type structure changes into a metallic phase with a hexagonal AlB$_2$-type structure upon partial substitution of Ni for Si in SrSi$_2$;  the hexagonal phase is stabilized at $0.1 < x <$ 0.7 in SrNi$_x$Si$_{2-x}$; superconductivity at up to 2.8 K appears at 0.1 $\leq$ $x$ $\leq$ 0.3.
The system showed a Pauli paramagnetic behavior, although the magnetic element Ni was substituted for Si.
Superconductivity, which was a conventional type-II type, emerged in the critical vicinity of the cubic-hexagonal phase boundary.

\section*{Acknowledgment}
We are indebted to Y. Ikeda, S. Araki, and T. C. Kobayashi for their help in using the arc furnace.
This work was partially performed at the Advanced Science Research Center, Okayama University. 
It was partially supported by Grants-in-Aids for Scientific Research from the Japan Society for the Promotion of Science (JSPS) and the Ministry of Education, Culture, Sports, Science, and Technology (MEXT), Japan.

\

\textit{Note added in proof} - We noticed a paper by K. Inoue, K. Kawashima, T. Ishikawa, M. Fukuma, T. Okumura, T. Muranaka, and J. Akimitsu, presented at 24th Int. Symposium on Superconductivity (ISS 2011) on October 26, 2011, reporting similar results.

\end{document}